# Fermi Surface of α-Uranium at Ambient Pressure

D. Graf,[1] R. Stillwell,[1] T.P. Murphy,[1] J.-H. Park,[1] M. Kano,[1] E.C. Palm,[1]
P. Schlottmann,[2] J. Bourg,[2] K.N. Collar,[2] J. Cooley,[3] J. Lashley,[3] J. Willit,[4] S.W. Tozer [1]

[1]National High Magnetic Field Laboratory, Florida State University, Tallahassee, Florida 32310, USA
[2]Department of Physics, Florida State University, Tallahassee, Florida 32306, USA
[3]Materials Science and Technology Division, Los Alamos National Laboratory, Los Alamos New Mexico 87545, USA
[4]Argonne National Laboratory, Argonne, Illinois 60439, USA

We have performed de Haas-van Alphen measurements of the Fermi surface of α-uranium single crystals at ambient pressure within the $\alpha_3$ charge density wave (CDW) state from 0.020 K - 10 K and magnetic fields to 35 T using torque magnetometry. The angular dependence of the resulting frequencies is described. Effective masses were measured and the Dingle temperature was determined to be 0.74 K ± 0.04 K. The observation of quantum oscillations within the $\alpha_3$ CDW state gives new insight into the effect of the charge density waves on the Fermi surface. In addition we observed no signature of superconductivity in either transport or magnetization down to 0.020 K indicating the possibility of a pressure-induced quantum critical point that separates the superconducting dome from the normal CDW phase.

71.18.+y  71.45.Lr  71.25.Pi

Uranium first isolated in the 1920's [1] has remained a challenge for condensed matter physics since the first scientific work on it began almost 90 years ago. The orthorhombic alpha phase of uranium (α-U) provides a unique setting to understand the role of *f*-electrons in the complex behavior of the actinides. The 20-atom unit cell of uranium and the three low temperature charge density waves are unique among the elements. These transitions, which are located at 43 K ($\alpha_1$), 37 K ($\alpha_2$) and 23 K ($\alpha_3$), result in the volume of the new unit cell below 23 K growing by a factor of 72 to ~6000 $A^3$ [1]. Uranium also undergoes high temperature phase transitions to the beta and gamma structures that render it impossible to grow high-quality single crystals of the alpha phase by the usual methods. In addition the crystal structure of α-U resembles corrugated cardboard along the [010] direction [2] makes the crystal particularly susceptible to twinning. Single crystals were serendipitously grown at Argonne National Laboratory while separating uranium from fission products in simulated spent uranium fuel [3]. These crystals proved to be of much higher quality than any previously available. Measurements on these crystals provided new insights into the physics of α-U [4, 5]. Although many outstanding questions remain including the origin of the CDW transitions, superconductivity and unique crystal structure, one issue is viewed as being a key to understanding the rest: the determination of the electronic structure of α-uranium. Here we report ambient pressure de Haas-van Alphen (dHvA) measurements of high quality crystals of α-U that have been enhanced with an annealing procedure to produce residual resistivity ratios (RRR) of up to 570, doubling the previous record [4]. Improved magnetic torque measurement techniques and DC fields to 35 T allowed the observation of 13 distinct frequencies for α-U, with effective masses less than or equal to 1.62 $m_e$. The quantum oscillations are observed after passing through the three CDW transitions shedding new light on the nature of the CDW's and their effect on the Fermi surface of α-U.

The fascinating physics of α-U is due to the complex behavior of the *f*-electrons. There is still much to understand regarding the *f*-electron behavior, and having experimental evidence of the electronic structure greatly aids modeling efforts. The *f*-orbitals at different sites overlap directly inducing a strongly metallic character in α-U. This results in the electrons being itinerant versus localized as found in Pu, Am and rare earth heavy fermions. Our ambient pressure measurements of the electronic structure of α-U are a key step in the experimental progress that will test theories and, in turn, provide predictive capabilities for the equation of state of this most complex element.

Charge density waves are quite common in a number of other materials e.g. $NbSe_3$[6], Cr[7], blue bronze[8], organic conductors[9] and $TaSe_2$[10]. These collective modes are usually associated with reduced dimensionality whereas U is a three dimensional metal with a low symmetry structure. In Cr the charge density wave appears in conjunction with a spin density wave over a small temperature interval [7]. The CDW's in α-U are a result of Peierls distortions, due to Fermi surface nesting of the complex band structure of the *f*-states [11]. The CDW's in α-U have historically made the observation of quantum oscillations quite difficult due to the concomitant stress and defects that were predicted to damage the crystal. [1, 4]. Because the observation of dHvA oscillations requires very clean crystals with long electron mean-free-paths, it was thought to be impossible without suppressing the CDW state via pressure. This was the motivation for the seminal high pressure dHvA measurements of Schirber and Arko [12].

Our measurements were carried out in a variety of systems including an Oxford top loading dilution refrigerator located in a 35 T resistive magnet, a Janis variable temperature insert (VTI) housed in a 31 T resistive magnet, a Quantum Design PPMS with a 16 T superconducting magnet and a Oxford top loading dilution refrigerator with a 20 T superconducting magnet. The α-U crystals were from a 1997 electrometallurgical growth using electrotransport through a LiCl-KCl eutectic flux containing 3% $UCl_3$ by weight [3]. The resistance as a function of temperature of an annealed sample is shown in Fig. 1 and the three CDW transitions are manifested in this data as steps at 43K, 37K and 23K as has been previously observed [4, 5]. They are more clearly seen in the dR/dT plot shown in the top panel of the figure. As seen in the inset to Fig. 1, the RRR is clearly enhanced from a typical unannealed value of 1-5 to a value of 570 indicating the low scattering rate in these crystals. Repeated thermal cycling from 300 K to 4 K did not degrade the RRR and the traces clearly showed hysteresis in the 23 K transition as observed in prior work [4]. The plateau at 180 K has been observed in several crystals, but a cause has not yet been determined. The crystals were oriented using both Laue x-ray backscatter and single crystal x-ray diffraction. Seiko piezo cantilevers were used to observe the quantum oscillations in all the magnet systems described above. A metal foil cantilever was also used to observe these oscillations using a dilution refrigerator in a 35 T resistive magnet and 20 T superconducting magnet. While the data was not of a comparable quality and is not reported here, these measurements did play the important role of confirming the results obtained using the piezo cantilevers.

Multiple crystals from multiple anneals were used for this study. All experiments below 1.5 K were performed with samples located in the dilute solution of the dilution refrigerators. From previous experience with self heating due to the radioactive decay of a $^{60}$Co single crystal of similar size and with a much higher total activity the resulting self heating in $\alpha$-U could only lead to a maximum temperature rise to 60 mK at the sample with the surrounding dilute mixture at 20 mK. Since the highest observed effective mass was 1.62 $m_e$ and the Dingle temperature was found to be ~ 0.74 K we conclude that radioactive self heating was not significant for our experiments.

Resistance and magnetization measurements to 20 mK showed no signs of the filamentary or bulk superconductivity previously found at ambient pressure between 0.1 K to 1.3 K.[1, 4, 13]. The application of pressure enhances $T_c$ and suppresses the CDWs leading to a maximum $T_c$ of 2.3 K at 1.1 GPa [1]. It has been posited that the superconductivity in $\alpha$-U is a result of impurity inclusions as $T_c$ is suppressed in cleaner samples [1, 4]. Given the large RRR and the lack of a superconducting transition in the present work the theory that superconductivity at ambient pressure in $\alpha$-U is impurity driven seems more likely. It is also possible that the internal crystal strain was low enough that the higher condensation energy of the CDW's, relative to that of superconductivity, completely blocked the formation of the superconducting state at even the lowest temperatures [1]. This possibility coupled with the results of our measurements agrees with the inability to observe bulk superconductivity at ambient pressure in prior measurements of $\alpha$-U [4]. The present scenario is then that $\alpha$-U has a pressure-induced superconducting dome. At low temperatures a pressure-tuned quantum critical point separates the normal phase from the superconducting phase. Further studies are underway to better understand the nature of superconductivity in alpha-uranium.

Figure 2 shows a representative torque signal from a magnetic field sweep with the background subtracted for a misalignment of 65 degrees from the [100] axis towards the [001] axis at 1.4 K. This angle was chosen because the observed torque signal becomes smaller as the field is aligned along one of the principal axes due to the alignment of the magnetization vector with the applied field. Oscillations are clearly visible (top panel) and a fast Fourier transform (FFT) analysis reveals multiple frequencies. The large amplitude of the quantum oscillations displayed further demonstrates the low scattering rate of the crystals used in this study.

In order to map out the Fermi surface of $\alpha$-U the orientation of the crystal was set via a rotation platform between [100] and [001] and also [010] to [001] and then the magnetic field was swept. Figures 3 & 4 show the angular dependence of the frequencies at various orientations. The intensity of the frequency peaks is indicated in the contour plots as color changes, while peak positions at selected angles are represented as solid black dots. The peak positions map the extremal area of the Fermi surface at ambient pressure. The observed Fermi surface has many more frequencies than those seen at high pressure by Schirber and Arko, especially at lower frequencies (e.g. F < 500 T). The origin could be small pockets due to incomplete Fermi surface nesting or the large magnetic fields available (i.e. 31 T) for our measurements which allowed 21 tesla of additional information beyond which was available to Schirber for determining the frequencies. A frequency common to both experiments is the surface described by the $F_1$ peak. This part of the Fermi surface is identical within the experimental uncertainty to the β orbit found by Schirber and Arko [12] showing that at least this portion of the Fermi surface is insensitive to applied pressure up to 0.8 GPa or the presence of the CDWs. The intensity of the $F_3$ frequency is stronger in the FFT analysis than that of the $F_1$ (β) frequency leading us to postulate that it is a result of the CDW's or the absence of applied pressure, as it was not observed by Schirber and Arko.

The differences in observed Fermi surfaces between this work and those of reference 12 show the role played by restructuring from the charge density waves, lattice changes from applied pressure, and the higher sensitivity of the measurement techniques used in this study. Given that the volume of the unit cell grows to 6000 Å$^3$ [1] in the α$_3$ state, it is not surprising that there are differences in the frequencies observed at ambient and high pressures. The element Cr also displays Fermi-surface changes from the SDW & CDW transitions, although they are from nesting incommensurate with the lattice rather than massive growth of the unit cell [7] due to Peierls distortions.

The effective masses for some of the measured frequencies were determined by the Lifshitz-Kosevitch relation from the temperature dependences of the frequency peaks[14]. These masses and frequencies are shown in Tables I, II and III. The measured effective masses (~ 1.6 $m_e$) are consistent with the Schirber and Arko work, as well as with measurements of the specific heat that determine γ = 9.13 mJ/(mol K$^2$) [5]and ARPES measurements [15]. The Dingle temperature was determined for the $F_1$, $F_3$ and $F_5$ peaks by measuring the quantum oscillations down to 0.020 K and is $T_D$ ~ 0.74 K for all three peaks. Effective masses for $F_6$-$F_{13}$ were not measured due to magnet time constraints.

In conclusion we observed a rich set of orbits for α-U in the ambient pressure α$_3$ charge density wave phase. These results show that the distortions of the lattice due to CDWs do not prevent the observation of quantum oscillations. This may be because in general materials that display CDWs are themselves very sensitive to strain which in the present case may have been reduced by the process of annealing the α-U crystals. This process, coupled with the increased sensitivity of the cantilever measurements, enabled the observation of dHvA oscillations in α-U at ambient pressure for the first time. These observations will enable accurate modeling of the behavior of the *f*-electrons in uranium and can be used as a check on band structure calculations. This discovery will allow the evolution of the Fermi surface to be mapped as the CDWs are suppressed with pressure. Superconductivity was not observed via resistivity or torque magnetometry down to

0.020 K in the crystals used in this study supporting the conjecture that superconductivity in α-U at ambient pressure is induced by defects and/or impurities. The present results in conjunction with previous ones lead to the conclusion that a pressure-induced quantum critical point separates the superconducting dome from the normal CDW phase. We hope that these results will stimulate further theoretical work on the Fermi surface of α-U.

The authors would like to thank Vaughn Williams and Robert Schwartz for technical support and Jim Schirber and Jim Smith for useful discussions and historical notes. Support for this work was provided by the DOE/NNSA under DE-FG52-06NA26193. This work was performed at the National High Magnetic Field Laboratory which is supported by NSF Cooperative Agreement No. DMR-0654118 and by the State of Florida. P.S. is supported by the DOE under. DE-FG02-98ER45707.

| Symbol | Frequency (T) | $m^*/m_0$ |
|---|---|---|
| $F_1$ | 1392 | 1.23 |
| $F_2$ | 1210 | 1.06 |
| $F_3$ | 740 | 1.33 |
| $F_4$ | 308 | 0.94 |
| $F_5$ | 95 | 0.68 |

Table I: Measured dHvA frequencies and effective masses with $H$ applied ~ 12 deg. off [100] rotating towards [001].

| Symbol | Frequency (T) | $m^*/m_0$ |
|---|---|---|
| $F_3$ | 692 | 1.62 |
| $F_4$ | 397 | 1.37 |
| $F_5$ | 85 | 0.59 |

Table II: Measured dHvA frequencies and effective masses with $H$ applied along [001].

| Symbol | Frequency (T) |
|---|---|
| $F_6$ | 1786 |
| $F_7$ | 1496 |
| $F_8$ | 1349 |
| $F_9$ | 1126 |
| $F_{10}$ | 448 |
| $F_{11}$ | 398 |
| $F_{12}$ | 331 |
| $F_{13}$ | 108 |

Table III. Measured dHvA frequencies with $H$ along [010]. $F_{11}$ is only visible starting from 36 deg. off of [010] rotating towards [001].

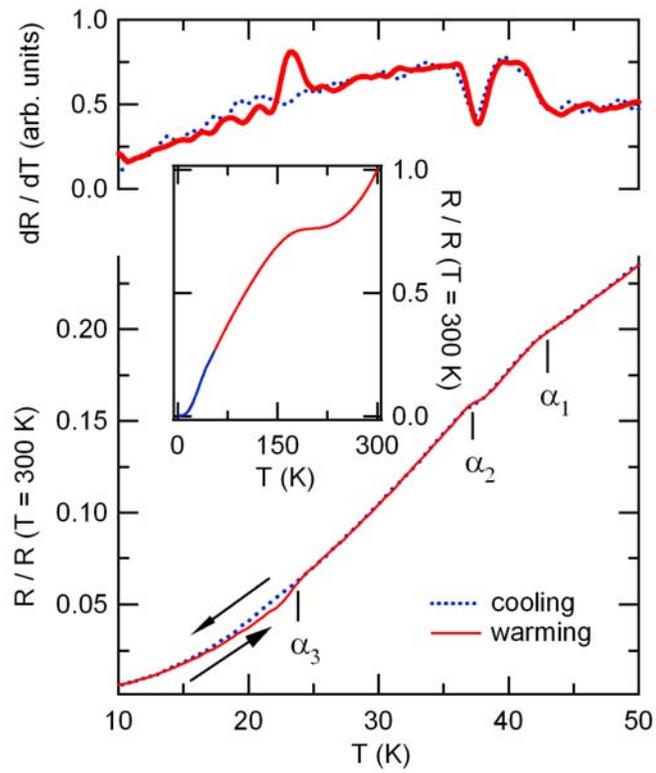

Figure 1. The zero-field resistivity and its derivative showing charge density wave transitions $\alpha_1$, $\alpha_2$ and $\alpha_3$ at 43K, 37K and 23K, respectively. Inset shows the overall resistivity change from room temperature to 1.8 K with RRR = 570.

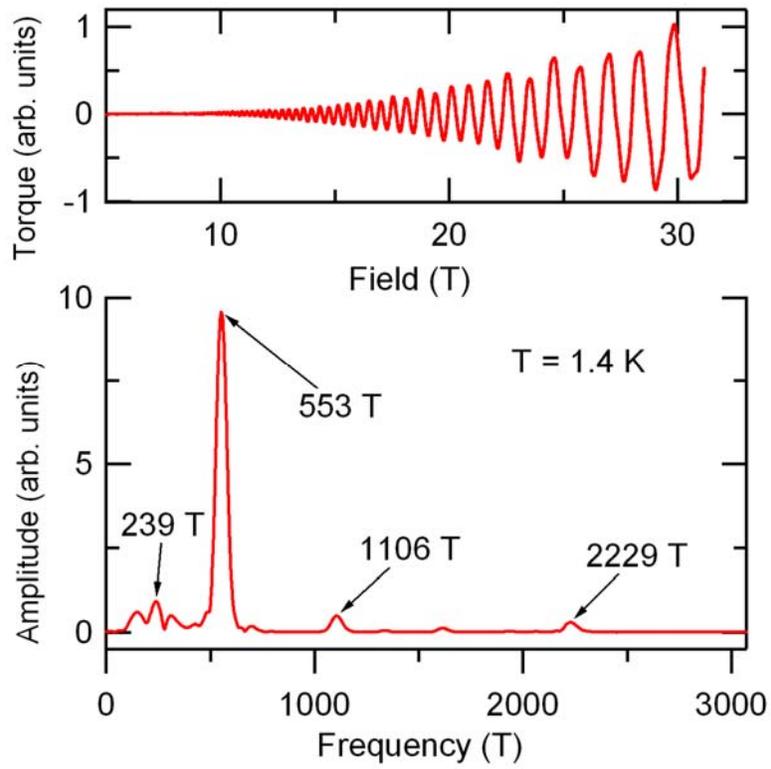

Figure 2. FFT result on background subtracted torque signal (top panel). The angle of the applied field is 65 deg. off [100] rotating towards [001]. From the angular dependence shown in Fig. 3 we can exclude the peak at 1106 T as a harmonic of 553 T.

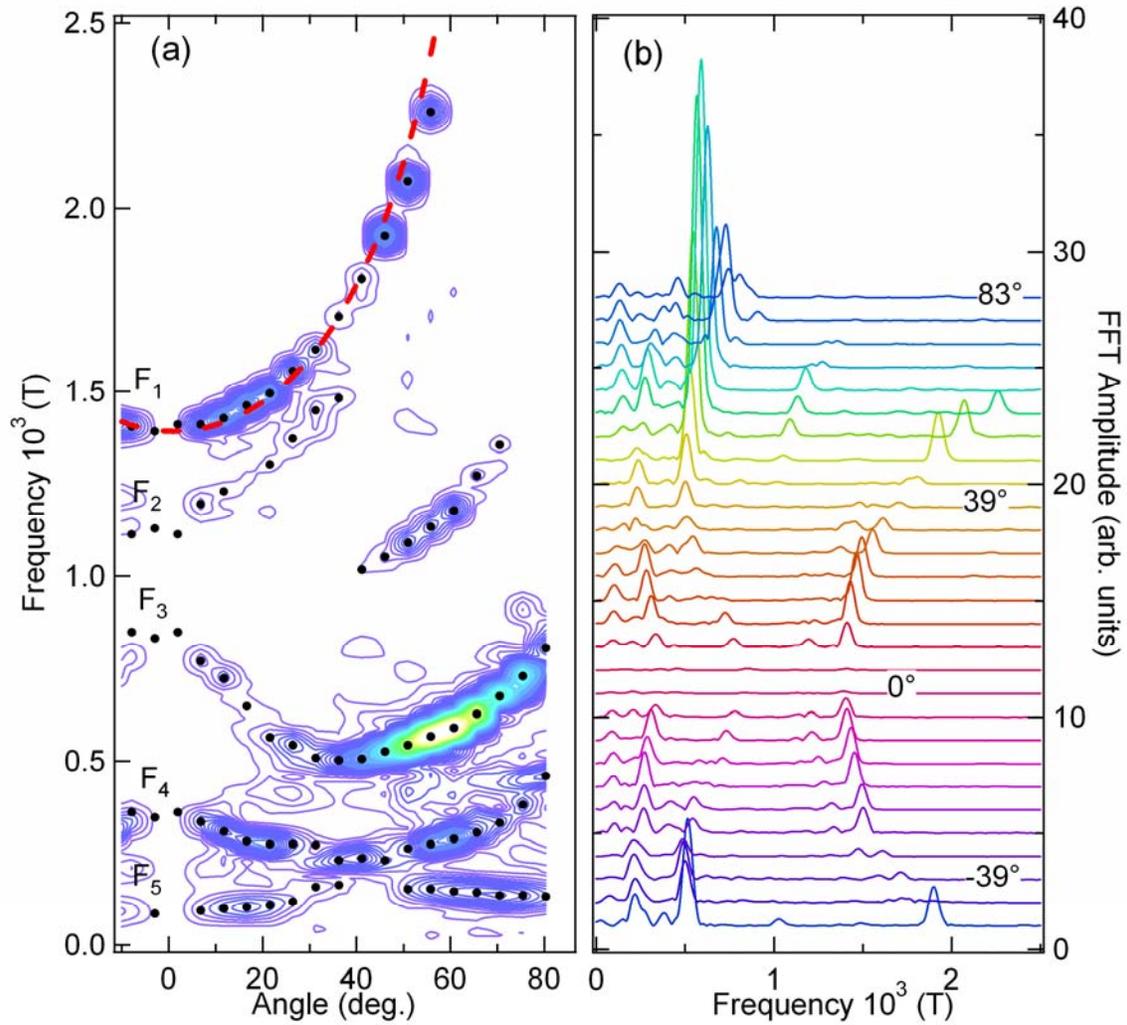

Figure 3. FFT results from background subtracted torque signal taken at T ~ 1.4 K and at various angles. Fig.3a is a contour plot with solid black dots at selected frequency peaks. Colors on the contour plot represent intensity of the FFT at fixed angle with yellow being most intense. The red dashed line represents a 1390/cos(θ) fit to the $F_1$ (β) orbit, in close agreement with the work of Schirber. Fig. 3b is a waterfall plot of the FFT for a series of angles. Offset is proportional to angular difference. Field sweeps at fixed angle were taken with the crystal rotated from [100] to [001] and the rotational axis parallel to [010].

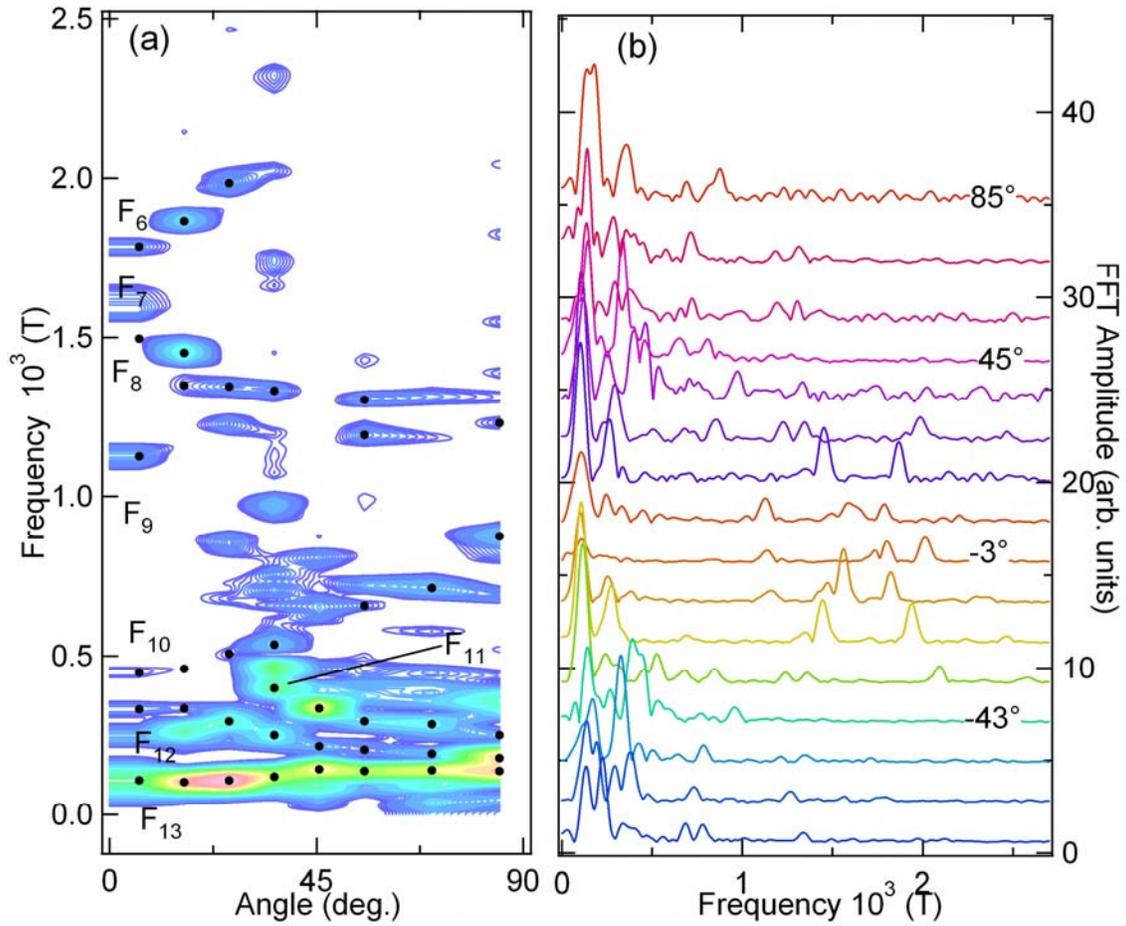

Figure 4. FFT results from background subtracted torque signal at angles between [010] and [001], with the rotational axis parallel to [100] and a temperature of 1.4K. Fig.4a is a contour plot with solid black dots at selected frequency peaks and the color red representing the highest intensity. Fig. 4b is a waterfall plot of the FFT for a series of angles. The offset is proportional to angular difference.

References:

[1] G. H. Lander, E. S. Fisher, and S. D. Bader, Adv. Phys. **43**, 1 (1994).
[2] L. T. Lloyd, and C. S. Barrett, Journal of Nuclear Materials **18**, 55 (1966).
[3] C. C. McPheeters *et al.*, Jom-Journal of the Minerals Metals & Materials Society **49**, 22 (1997).
[4] G. M. Schmiedeshoff *et al.*, Philos. Mag. **84**, 2001 (2004).
[5] J. C. Lashley *et al.*, Physical Review B **63**, 224510 (2001).
[6] M. Ido *et al.*, Journal of the Physical Society of Japan **59**, 1341 (1990).
[7] E. Fawcett, Rev. Mod. Phys. **60**, 209 (1988).
[8] J. P. Pouget *et al.*, Journal De Physique **46**, 1731 (1985).
[9] D. Jerome, and H. J. Schulz, Adv. Phys. **51**, 293 (2002).
[10] D. E. Moncton, J. D. Axe, and F. J. Disalvo, Physical Review B **16**, 801 (1977).
[11] L. Fast *et al.*, Physical Review Letters **81**, 2978 (1998).
[12] J. E. Schirber, and A. J. Arko, Physical Review B **21**, 2175 (1980).
[13] J. L. O'Brien *et al.*, Physical Review B **66**, 064523 (2002).
[14] D. Shoenberg, *Magnetic Oscillations in Metals* (Cambridge University Press, New York, 1984).
[15] C. P. Opeil *et al.*, Physical Review B **75**, 045120 (2007).